\documentclass[aps,prd,twocolumn,superscriptaddress,nofootinbib,10pt]{revtex4-1}
\usepackage[paperwidth=21cm,paperheight=29.7cm,top=2.54cm,bottom=2.54cm,left=2cm,right=2cm]{geometry}
\usepackage[colorlinks,linkcolor=blue,anchorcolor=blue,citecolor=blue,urlcolor=blue]{hyperref}
\usepackage{graphicx,subfigure}
\usepackage[figuresright]{rotating}
\usepackage{amsmath,amsfonts,amssymb,bm}
\usepackage{array,enumitem,multirow}
\usepackage{acronym}
\usepackage{makecell}
\newcommand{\be}{\begin{equation}}
\newcommand{\ee}{\end{equation}}
\newcommand{\bea}{\begin{eqnarray}}
\newcommand{\eea}{\end{eqnarray}}

\newcommand{\SOUTHCUT}{
School of Physics and Optoelectronics, South China University of Technology, Guangzhou 510641,
People's Republic of China}
\newcommand{\bk}{\mathsf{k}}
\def\mc{\mathcal}

\newacro{GR}{general relativity}
\newacro{GW}{gravitational wave}
\newacro{MG}{modified gravity theory}
\newacro{BH}{Black hole}
\newacro{PN}{post-Newtonion}
\newacro{ppE}{parameterized post-Einsteinian}
\newacro{GCB}{galactic ultra-compact binary}
\newacro{SBHB}{stellar-mass black hole binary}
\newacro{MBHB}{massive black hole binary}
\newacro{BHB}{black hole binary}
\newacro{IMBHB}{intermediate-mass black hole binary}
\newacro{EMRI}{extreme mass ratio inspiral}
\newacro{IMRI}{intermediate mass ratio inspiral}
\newacro{SGWB}{stochastic gravitational wave background}
\newacro{MECO}{minimal energy circular orbit}
\newacro{FAR}{false alarm rate}
\newacro{CE}{Cosmic Explorer}
\newacro{ET}{Einstein Telescope}
\newacro{LISA}{Laser Interferometer Space Antenna}
\newacro{EdGB}{Einstein-dilaton Gauss-Bonnet}
\newacro{dCS}{dynamic Chern-Simons}
\newacro{SNR}{signal-to-noise ratio}
\newacro{FIM}{Fisher Information Matrix}
\newacro{ISCO}{innermost stable circular orbit}
\newacro{NSBH}{neutron star-black hole binary}
\newacro{MCMC}{Markov Chain Monte Carlo}
\newacro{QNM}{quasi-nomral mode}

\begin{document}

\title{Probing the Tidal Deformability of the Central Object with Analytic Kludge Waveforms of an Extreme Mass Ratio Inspiral }

\author{Tieguang Zi}
\email{zitg@scut.edu.cn}
\author{Peng-Cheng Li}
\email{pchli2021@scut.edu.cn, corresponding author}

\affiliation{\SOUTHCUT}

\date{\today}

\begin{abstract}
We  develop approximate ``analytic-kludge" waveforms to describe the inspiral of a stellar-mass compact object into a supermassive compact object in an extreme mass ratio inspiral (EMRI) scenario. The deformability of the supermassive compact object is characterized by a dimensionless quantity called the tidal Love number (TLN). Our analysis shows that, up to the leading order of the mass ratio, the conservative dynamics of the EMRI are not affected by tidal interaction, and the tidal effect is only present in the induced quadrupole moment. We calculate the energy and angular momentum fluxes and obtain leading order corrections to the orbital evolution equations. By comparing the waveforms with and without tidal interaction, we demonstrate that even a small TLN can produce significant differences in the waveforms, which can be detected by space-borne detector LISA. Finally, using the Fisher information matrix method, we perform parameter estimation for the TLN and find that the precision can reach the level of $10^{-4}$ in suitable scenarios. 
\end{abstract}

\maketitle

\section{Introduction}

The inspiral of stellar-mass compact objects (COs) into supermassive compact objects (SMCOs) at the center of galaxies presents an especially interesting gravitational-wave (GW) source for future space-based GW detectors such as LISA \cite{LISA:2017pwj}, TainQin \cite{TianQin:2020hid}, and Taiji \cite{Hu:2017mde}. These events are commonly known as extreme-mass-ratio inspirals (EMRIs) since the ratio of CO mass to SMCO mass is typically $10^{-4}-10^{-7}$. Due to the emission of GWs and the extreme mass ratio, these systems inspiral slowly, completing $10^4-10^5$ cycles in the frequency band of those space-based detectors. As a result, the GW signals from EMRIs contain a wealth of information about the surroundings of the SMCOs. Detection of these signals would not only help answer key astrophysical questions \cite{Berry:2019wgg, Amaro-Seoane:2022rxf}, but also provide new insights into fundamental physics, such as tests of general relativity (GR), the nature of black holes (BHs), and more \cite{Babak:2017tow,Barausse:2020rsu,Zi:2021pdp,LISA:2022kgy}.

An efficient way to distinguish between BHs and exotic compact objects (ECOs) \cite{Cardoso:2019rvt} and test GR is to measure the effect of tidal deformability on the GWs emitted by compact binaries. In a compact binary, each object experiences a tidal field generated by the gravitational field of its companion, which modifies the dynamical evolution of the system and the GW emission \cite{Flanagan:2007ix}. In the adiabatic limit, the imprint of the tidal interaction on the GW waveform is encoded by the tidal Love numbers (TLNs)\cite{Poisson:2014book}, which are constant quantities sensitive to the internal structure of the object. So far, TLN measurements have successfully constrained the equation of state of neutron stars \cite{Hinderer:2007mb,LIGOScientific:2018cki}.

A crucial fact concerning GW observations is that the TLNs of a BH in GR are precisely zero. This was first shown for Schwarzschild BHs \cite{Binnington:2009bb,Damour:2009vw,Gurlebeck:2015xpa}, the same result was then shown to apply to Kerr BHs with slow rotation \cite{Landry:2015zfa,Pani:2015nua,Poisson:2014gka} and finally with arbitrary spin by different groups \cite{Charalambous:2021kcz,Charalambous:2021mea,Chia:2020yla,Hui:2020xxx,LeTiec:2020spy,LeTiec:2020bos}.
However, the TLNs are generically not zero for ECOs and for BHs in gravities alterative to GR \cite{Cardoso:2017cfl,Cardoso:2019rvt,Nair:2022xfm,DeLuca:2022tkm}.
Thus, if one measures a nonvanishing TLN in GWs from compact binaries, which may indicate the existence of ECO or the deviation of GR. At present, the measurements of the TLNs have been employed to analyze the GW events observed by LIGO and Virgo \cite{Johnson-Mcdaniel:2018cdu,Narikawa:2021pak}, and the measurement capability by LISA for comparable-mass binaries has also been studied \cite{Cardoso:2017cfl,Maselli:2017cmm}.

Recently, by working within the post-Newtonian (PN) approximation, Pani et al.\cite{Pani:2019cyc} (and also \cite{Datta:2021hvm}) found that the TLN of the central object of EMRIs affects the gravitational waveform at the leading order of the mass ratio, which has the equal contribution to the phase as the ordinary radiation-reaction term. This means the space-based GW detectors such as LISA could place very stringent constraints on the TLNs of the central object. Furthermore, ref. \cite{Piovano:2022ojl} conducted a more in-depth analysis for the estimation of the measurement of the tidal deformability of a SMCO through an EMRI detection by LISA. The authors considered a hybrid ``Teukolsky+PN'' waveform where tidal corrections to the energy flux are introduced with their corresponding PN terms and the authors found the TLN of the central SMCO can be measured at the level of $10^{-3}$ if the central object is highly spinning.

It is well-established that accurately detecting and analyzing extreme mass-ratio inspirals (EMRIs) requires the construction of precise waveform models, which are typically generated using strong-field perturbation theory. However, such models can be computationally expensive \cite{Pound:2021qin,Chua:2020stf,Katz:2021yft}. To expedite the process, many EMRI parameter estimation studies utilize ``kludge" models \cite{Barack:2003fp,Babak:2006uv,Chua:2017ujo}. The first kludge model proposed by Barack and Cutler is known as the ``analytic kludge" (AK) model. In this model, the compact object moves in a quasi-Keplerian ellipse, with its orbital parameters slowly evolving due to radiation reaction. The waveform is then generated using the well-known Peter-Mathews formula under the quadrupole approximation \cite{Peters:1963ux,Peters:1964zz}. Although the calculation is done under the PN approximation, the AK model can still capture important features of accurate EMRI waveforms, including the relativistic precession of the orbital plane and pericenter.

In this paper, we would like to study the tidal deformability of the SMCO of an EMRI within the framework of AK model \cite{Barack:2003fp}. We will study how the tidal deformability of the SMCO caused by the CO modifies the  evolution equations of various orbital parameters. As we will show that in the extreme-mass-ratio case, this is reflected only in the modification to the fluxes of energy and angular momentum of the gravitational radiation. Furthermore, to
quantify the effects of the TLN on the waveforms, we will compute the mismatches between waveforms from EMRIs with and without the tidal interaction. Finally, we will perform parameter estimation of the TLN for the SMCO with space-borne GW detectors LISA using the Fisher information matrix method.

The paper is organized as follows. In Sec. \ref{EMRIwaveforms}, we present the derivation of the modified AK waveforms when taking the tidal interaction between the CO and SMCO into account. In Sec. \ref{results}, we study the comparison of waveforms with and without the tidal interaction and obtain the constraint on the TLN of the SMCO through the detection of the EMRIs by LISA. Finally, we give a brief summary in
Sec. \ref{summary}. The details of the Fourier decomposition of the tidal-induced inertial tensor is given in Appendix \ref{appA}. Throughout this paper, we use the geometric units,
where $c=1=G$.
\section{EMRI waveforms}\label{EMRIwaveforms}
\subsection{Conservative dynamics in the adiabatic limit}
For an EMRI system  consists of a CO with mass $m$ and a SMCO with mass $M$, satisfying $M\gg m$, up to the leading order of the mass ratio $q=m/M$, we have the total mass $m_{tot}=M+m\simeq M$, the reduced mass $\mu=\frac{m M}{m+M}\simeq m$ and the symmetric mass ratio $\eta=mM/(m+M)^2\simeq q$. According to the analysis in \cite{Pani:2019cyc}, due to the extreme mass ratio only the TLN of the central object of the EMRI affects the waveform and the one of the CO can be neglected.  Thus, in this work we only consider the SMCO is deformable.

In Newtonian gravity,  the tidal field felt by the SMCO is characterized by the tidal moment, which is defined as
coefficients in the Taylor expansion of the external potential about the center-of-mass position \cite{Vines:2010ca}. Up to quadrupole order we have
the tidal moment
\be
G^{ij}_2=-\partial_i \partial_j U_{ext}=\frac{3 m}{r^3}\left(n^in^j-\frac{1}{3}\delta^{ij}\right),
\ee
where $U_{ext}$ is the external potential felt by the SMCO and is sourced by the CO. Moreover, $x^i$ is the relative position vector
between the SMCO and the CO, $r=\sqrt{\delta_{ij}x^ix^j}$ and $n^i=x^i/r$. The tidal deformation of the SMCO is described at leading order by the mass quadrupole moment
\be
Q^{ij}_2=\int d^3 y \rho (y^i y^j-\frac{1}{3} \delta^{ij} y^ky_k),
\ee
where $\rho(t,y^i)$ is the mass density and $y^i$ is the displacement from the SMCO's center-of-mass position.
In the absence of the non-uniform gravitational field from the companion, viz., the CO, the SMCO would be  spherical and its quadrupole moment would vanish. In the adiabatic limit, when the response
time scale of the SMCO is much less than the time scale
on which the tidal field changes, the induced quadrupole moment
will be given \cite{Vines:2010ca}
\be\label{tidalquadrupole}
Q^{ij}_2=\lambda G^{ij}_2,
\ee
where the constant $\lambda$ is called the tidal deformability.
This is related to a dimensionless constant by \cite{Pani:2019cyc} ,
\be\label{1ambdaTLN}
\lambda=\frac{2}{3}M^5\bk,
\ee
where $\bk$ is the well-known TLN \footnote{The TLN is more often defined by $\lambda=\frac{2}{3}R^5\bk$, where $R$ is the body's radius \cite{Mora:2003wt}.}.

Working in the center-of-mass frame and up to the quadrupole-tidal interaction, the Lagrangian describing the evolution of the EMRI is given by
\be
\mc{L}=\frac{\mu v^2}{2}+\frac{\mu M}{r}-U_Q+\mathcal{L}_2^{int},
\ee
where $v^2=\delta_{ij}\dot{x}^i\dot{x}^j$ with dot denoting derivatives with respect the coordinate time $t$, $U_Q$ is the potential energy of the quadrupole-tidal interaction
\be
U_Q=-\frac{1}{2}Q^{ij}_2G^{ij}_2,
\ee
and $\mathcal{L}_2^{int}$ is the internal Lagrangian for the SMCO, which in the adiabatic limit can be taken as
\be
\mathcal{L}_2^{int}=-\frac{1}{4\lambda}Q^{ij}_2Q^{ij}_2.
\ee
Plugging Eq. (\ref{tidalquadrupole}) into above expressions and from the Euler-Lagrangian equation, we obtain the orbital equation of motion
\be
\ddot{x}^i=-\frac{M n^i}{r^2}\left(1+\frac{9}{r^5}\lambda q\right).
\ee
The orbital equation of motion admits circular orbits as solutions. However, the more general quasi-Keplerian orbits are needed for the EMRIs. To our knowledge, this problem has only been addressed to some extent \cite{Bernaldez:2023xoh}.
The second term on the right hand side of above equation can be treated as a perturbative term, since the tidal force is in general weaker than the Newtonian force, and more importantly, because the strength of the second term is suppressed by the mass ratio $q$. This fact allows us to employ the method of osculating orbital elements \cite{Poisson:2014book} to solve the problem of perturbed Keplerian orbits.
The basic ideal behind this method is that there always exists a Keplerian orbit with time-dependent orbital elements that is tangent to the perturbed orbit at that time. Thus, we can still write the distance between the SMCO and the CO as
\be\label{rpsi}
r=\frac{p}{1+e  \cos \psi},
\ee
where $p$ is the semi-latus rectum, $e$ is the eccentricity and $\psi$ is the true anomaly of the orbits. However, in general both $p$ and $e$ and other orbital elements are functions of time and not constants anymore.
Since the osculating equations for a general perturbative force can be found in \cite{Poisson:2014book}, here we directly apply them to our problem where the tidal force is along the radial direction. Then the osculating equations are given by
\be\label{tidalforcep}
\frac{dp}{dt}=0,
\ee
\be\label{tidalforcee}
\frac{de}{dt}=- \frac{9q\lambda}{r^7}\sqrt{pM} \sin\psi  ,
\ee
\be
\frac{d\omega}{dt}=\frac{9q\lambda}{r^7}\frac{\sqrt{pM}}{e} \cos \psi ,
\ee
\be\label{tidalforcepsi}
\frac{d\psi}{dt}=\sqrt{\frac{M}{p^3}}(1+e\cos \psi)^2-\frac{9q\lambda}{r^7}\frac{\sqrt{pM} }{e}\cos \psi,
\ee
where $\omega$ is  the longitude of pericenter defined specifically as the angle between the line of nodes and the
direction to the pericenter, as measured in the orbital plane. So the tidal force will cause the precession
of the pericenter.

Due to the presence of tidal terms, the last three equations must be solved numerically, which makes it difficult to calculate the energy and angular momentum fluxes of GWs. However, we can observe that the corrections resulting from tidal terms are proportional to the mass ratio $q$. This means that the contribution of these corrections to the energy and angular momentum fluxes of GWs can be neglected. Consequently, both the semi-latus rectum and the eccentricity can be treated as constants during flux calculations.
Alternatively, since the corrections are heavily suppressed by the mass ratio, they act on a much longer timescale than the orbital period, similar to the case of radiation reaction. Thus, we can compute the average values of $\dot{e}$ and $\dot{\omega}$ over the orbital period, which are known as secular changes. At leading order of the mass ratio, the average of $\dot{e}$ is zero. In contrast, the secular change of  $\dot{\omega}$ is not vanishing. This precession phenomenon is called apsidal advance in astronomy \cite{Poisson:2014book}. However, the precession
of the pericenter caused by relativistic effect is of the order $O(q^0)$ so is dominant than the apsidal advance. Therefore, we can conclude that, up to leading order of the mass ratio, the tidal interaction between the SMCO and the CO does not affect the conservative dynamics of EMRIs.

The method of osculating orbital elements allows the orbital energy and the angular momentum to have the same form as in the Keplerian case, thus
\be\label{orbitalenergy}
E=-\frac{\mu M}{2p}(1-e^2),
\ee
and
\be\label{orbitalanagularmomentum}
L_z=\mu \sqrt{M p}.
\ee
The osculating equations reveal that the tidal force has no effect on the orbital momentum, but it does affect the orbital energy through the eccentricity of the orbit. Thus, in the presence of the tidal interaction, the orbital angular momentum remains conserved while the orbital energy is not. However, since the tidal corrections are suppressed by the mass ratio, the orbital energy is the same as the Keplerian one up to leading order of the mass ratio.
\subsection{Fluxes}
Now we consider the dissipative dynamics of the EMRIs in the presence of tidal interaction. We would like to calculate the change rates of the eccentricity $e$ and the radial orbital frequency $\nu$ with respect to the coordinate time, due to the energy flux and the angular momentum flux from the gravitational radiation.

For the gravitational radiation, the standard
quadrupole formulas of the energy flux and the angular momentum flux are given by \cite{Peters:1963ux,Peters:1964zz}
\be
\frac{dE}{dt}=\frac{1}{5}\left<\frac{d^3Q_{ij}}{dt^3} \frac{d^3Q^{ij}}{dt^3}\right>,
\ee
and
\be
\frac{dL_i}{dt}=\frac{2}{5}\epsilon_{ijk}\left<\frac{d^2Q_{jm}}{dt^2} \frac{d^3Q^{km}}{dt^3}\right>,
\ee
where the quadrupole moment is now given by \cite{Vines:2011ud}
\be\label{quadrupolemoment}
Q^{ij}=\mu r^2 \left(n^i n^j-\frac{1}{3}\delta^{ij}\right)+Q_2^{ij},
\ee
where $Q_2^{ij}$ is the quadrupole moment Eq. (\ref{tidalquadrupole}) induced by the tidal field.
Besides, the angle-brackets denote the average over one cyclic
motion in $r$, which via Eq. (\ref{rpsi}) can be turned into the integral for $\psi$, e.g.,
\be
\left<X\right>=\frac{1}{T}\int^{T}_0X(t)dt=\frac{1}{T}\int^{2\pi}_0X(\psi)\frac{d\psi}{\dot{\psi}},
\ee
where the period $T$ is given by
\be
T=\int^{2\pi}_0\frac{d\psi}{\dot{\psi}}.
\ee
To perform the calculation of the energy and angular momentum fluxes, we should first notice that
the induced quadrupole moment in Eq. (\ref{quadrupolemoment}) is not suppressed by the mass ratio. This can be seen as follows:
\be
\frac{\lambda\frac{3 m}{r^3}}{\mu r^2}=2\bk \left(\frac{M}{r}\right)^5 \frac{m}{\mu}\simeq 2\bk \left(\frac{M}{r}\right)^5.
\ee
As a result, when computing the derivatives of the quadrupole moment with respect to time, the contribution from the tidal correction in Eq. (\ref{tidalforcee}) will be suppressed by the mass ratio. Moreover, when averaging over the orbital period, the influence of the second term in Eq. (\ref{tidalforcepsi}) will also be suppressed by the mass ratio. Therefore, in the calculation of energy and angular momentum fluxes, the orbits can be approximated as Keplerian orbits, with the effect of tidal interaction encoded only in the induced quadrupole moment. This significant simplification arises from the tiny mass ratio of the EMRIs and will not occur for inspirals of binaries with comparable masses.

Direct calculations lead to
\bea\label{GWdEdt}
\frac{dE}{dt}=f_1(e)\left(\frac{M}{p}\right)^5q^2+f_2(e) \left(\frac{M}{p}\right)^{10}\bk  q^2,\nonumber\\
\eea
\bea\label{GWdELzdt}
\frac{dL_z}{dt}=g_1(e)M\left(\frac{M}{p}\right)^{7/2}q ^2+g_2(e)M\left(\frac{M}{p}\right)^{17/2}\bk q^2,\nonumber\\
\eea
where the related coefficients are all functions of the eccentricity only
\be
f_1(e)=\frac{\left(1-e^2\right)^{3/2}  }{15 } \left(37 e^4+292 e^2+96\right),
\ee
\bea
f_2(e)&=&\frac{ \left(1-e^2\right)^{3/2}   }{20 }(225 e^{10}+10355 e^8\\
&&+50200 e^6+53904 e^4+13504 e^2+512) ,\nonumber
\eea
\be
g_1(e)=\frac{4 \left(1-e^2\right)^{3/2}   }{5 }\left(7 e^2+8\right),
\ee
and
\bea
g_2(e)&=&\frac{ \left(1-e^2\right)^{3/2}   }{20 }(165 e^8+5080 e^6\nonumber\\
&&+14640 e^4+7488 e^2+512).
\eea
One can observe that $f_1(e)$ and $g_1(e)$ match the results in the case without the tidal interaction \cite{Peters:1963ux,Peters:1964zz}. Additionally, when $e=0$ and the leading order of the mass ratio is retained, the results are the same as the Newtonian ones presented in \cite{Henry:2019xhg,Henry:2020ski}.

From Eqs. (\ref{orbitalenergy}) and (\ref{orbitalanagularmomentum}), we can obtain the rates of change in the orbital energy and angular momentum with respect to time,
\be
\frac{dE}{dt}=\frac{\mu  M e }{p}\frac{de}{dt}+\frac{\mu  M \left(1-e^2\right)}{2 p^2}\frac{dp}{dt},
\ee
and
\be
\frac{dL_z}{dt}=\frac{\mu}{2}\sqrt{\frac{M}{p}}\frac{dp}{dt}.
\ee
Due to the balance condition, the gravitational radiation will cause the loss of the orbital energy and angular momentum, as a consequence both $p$ and $e$ will decay with the coordinate time. Combine above two equations with
Eqs. (\ref{GWdEdt}) and (\ref{GWdELzdt}), we obtain
\bea
\frac{dp}{dt}=-2 g_1(e)\left(\frac{M}{p}\right)^{3}q-2g_2(e)\left(\frac{M}{p}\right)^{8}\bk q,
\eea
and
\bea
\frac{de}{dt}&=&-\frac{q}{M e}\left[f_1(e)-(1-e^2)g_1(e)\right]\left(\frac{M}{p}\right)^4\nonumber\\
&&-\frac{\bk q}{M e} \left[f_2(e)-(1-e^2)g_2(e)\right]\left(\frac{M}{p}\right)^{9}.
\eea
Remember that $e$ is also affected by the tidal force, so we should combine these equations with Eq. (\ref{tidalforcee}) since the contribution from the tidal force occurs at linear order of $q$ as well, then we have
\bea
\frac{de}{dt}&=&-\frac{q}{M e}\left[f_1(e)-(1-e^2)g_1(e)\right]\left(\frac{M}{p}\right)^4\nonumber\\
&&
-\frac{\bk q}{M e} \left[f_2(e)-(1-e^2)g_2(e)\right]\left(\frac{M}{p}\right)^{9}\nonumber\\
&&-\frac{6\bk q}{M}\sin\psi (1+e\cos\psi)^7\left(\frac{M}{p}\right)^{13/2}.
\eea
For a Keplerian orbit, it is often express the semi-latus rectum $p$ with the radial orbital frequency, which has a linear connection with the change rate of the mean anomaly to time. Due to the method of osculating orbital elements, similar to the Kepler's third law, for the perturbed orbits the semi-latus rectum can still be written as 
\be
p=\frac{M(1-e^2)}{(2\pi M\nu)^{2/3}},
\ee
where $\nu$ is the radial orbital frequency. Then we can obtain
\bea
\frac{d\nu}{dt}&=&\frac{ 3q }{2 \pi   M^2}(2\pi M\nu)^{11/3}(1-e^2)^{-5}f_1(e)\nonumber\\
&&+\frac{3\bk q }{2\pi   M^2}(2\pi M\nu)^7\left(1-e^2\right)^{-10}f_2(e)\\
&&+\frac{9 \bk q e}{ \pi   M^2}\sin\psi (1+e\cos\psi)^7(2\pi M\nu)^{16/3}\left(1-e^2\right)^{-15/2},\nonumber
\eea
and
\bea
\frac{de}{dt}&=&-\frac{q}{M e}\left[f_1(e)-(1-e^2)g_1(e)\right]\left(2\pi M\nu\right)^{8/3}\left(1-e^2\right)^{-4}\nonumber\\
&&-\frac{\bk q}{M e} \left[f_2(e)-(1-e^2)g_2(e)\right] \left(2\pi M\nu\right)^{6}  \left(1-e^2\right)^{-9} \nonumber\\
&&-\frac{6\bk q}{M}\sin\psi (1+e\cos\psi)^7\left(2\pi M\nu\right)^{13/3}\left(1-e^2\right)^{-13/2}.\nonumber\\
\eea
Obviously, the last terms in the two equations above stem from  the effect of the tidal force on the conservative dynamics Eq. (\ref{tidalforcee}). They are indeed of the same order as the results from  the radiation reaction, with both appearing at the linear order of the mass ratio. Therefore, we can also perform the average over the period time as we have done for the energy and the angular momentum fluxes. A simple calculation shows that averages of the last terms in the two equations above are zero.
\subsection{AK waveforms}
In this subsection, we provide a brief review of the AK waveforms \cite{Barack:2003fp} and the necessary modifications due to the presence of the tidal interaction. In the AK model, EMRIs are approximated as a Keplerian binary at any given time emitting a lowest order, quadrupole waveform. Furthermore, the orbital parameters are governed by PN equations, which include orbital decay from radiation reaction, pericenter precession, and Lense-Thirring precession of the orbital plane.

In the previous subsection, we have obtained the leading order equations describing the evolution of the radial orbital frequency and the eccentricity in the presence of the tidal interaction. We combine these leading order
corrected equations with those higher-order PN equations
in the original AK model. Then the complete orbital
evolution equations are given by
\be
\dot{\Phi}=2\pi\nu,
\ee
\bea
\dot{\nu}&=&\frac{ 3q }{2 \pi   M^2}(2\pi M\nu)^{11/3}(1-e^2)^{-5}f_1(e)\nonumber\\
&&+\frac{3\bk q }{2\pi   M^2}(2\pi M\nu)^7\left(1-e^2\right)^{-10}f_2(e)\nonumber\\
&&+\left(\frac{1273}{336}-\frac{2561}{224}e^2-\frac{3885}{128}e^4-\frac{13147}{5376}e^6\right)\nonumber\\
&&\times(2\pi M\nu)^{2/3}\nonumber\\
&&-(2\pi M\nu)a \cos\lambda(1-e^2)^{-1/2}\Big(\frac{73}{12}+\frac{1211}{24}e^2\nonumber\\
&&+\frac{3143}{96}e^4+\frac{65}{64}e^6\Big),
\eea
\bea
\dot{e}&=&-\frac{q}{M e}\left[f_1(e)-(1-e^2)g_1(e)\right]\left(2\pi M\nu\right)^{8/3}\left(1-e^2\right)^{-4}\nonumber\\
&&-\frac{\bk q}{M e} \left[f_2(e)-(1-e^2)g_2(e)\right] \left(2\pi M\nu\right)^{6}  \left(1-e^2\right)^{-9} \nonumber\\
&&-\frac{1}{56}(2\pi M\nu)^{2/3}(133640+108984e^2-25211e^4)\Big]
\nonumber\\
&&+e \frac{q}{M}a \cos\lambda (2\pi M\nu)^{11/3}(1-e^2)^{-4}\nonumber\\
&&\times\left(\frac{1364}{5}+\frac{5032}{15}e^2+\frac{263}{10}e^4\right)
\eea
\be
\dot{\alpha}~=~ \frac{2a}{M}(2\pi M\nu)^2\left(1-e^2\right)^{-3/2},
\ee
\bea
\dot{\tilde{\gamma}}&=& 6\pi \nu (2\pi M\nu)^{2/3}\left(1-e^2\right)^{-1}\nonumber\\
&&\times\Big[1+\frac{1}{4}(2\pi M\nu)^{2/3}\left(1-e^2\right)^{-1}(26-15e^2)\Big]\nonumber\\
 &&- 12\pi \nu a \cos\lambda (2\pi M\nu) \left(1-e^2\right)^{-3/2}.\label{gammadot}
\eea
The equation for $\dot{\nu}$ and $\dot{e}$ are given accurately through $3.5$ PN order, the equations for $\dot{\tilde{\gamma}}$ and  $\dot{\alpha}$ are accurate through $2$ PN order.
Here $\Phi$ is known as the mean anomaly, $\lambda$ is the inclination angle of the
orbital plane with respect to the spin direction of the SMCO and $a$ is the dimensionless spin parameter of the SMCO.
Moreover, $\alpha $ is the azimuthal direction of the orbital angular momentum in the spin-equatorial plane and $\tilde{\gamma}$ is the angle between
$\hat{L}\times\hat{S}$ and pericenter, where $\hat{L}$ is the unit vector of the orbital angular momentum and $\hat{S}$ is the unit vector of the SMCO's spin. So $\dot{\alpha}$ describes the Lense-Thirring precession of the orbital plane and $\dot{\tilde{\gamma}}$ describes the pericenter precession. From Eq.(\ref{gammadot}) we can see that although the tidal force can cause the precession of the pericenter, the effect only appears at the linear order of the mass ratio, which is significantly suppressed when compared with the relativistic precession of the pericenter.

To work within the framework of Barack and Cutler \cite{Barack:2003fp}, where the orbital evolution equations involve the mean anomaly instead of the true ananomy, in the following we need to study the Fourier decomposition of the quadrupole
radiation in the presence of the tidal interaction. In the quadrupole approximation and taking the transverse and traceless gauge, the GW strain in the weak field regime is given by
\be
h_{ij}=\frac{2}{D}\left(P_{ik}P_{jl}-\frac{1}{2}P_{ij}P_{kl}\right)\ddot{\mathbb{I}}^{kl},
\ee
where $D$ is the distance to the source, $P_{ij}=\delta_{ij}-\hat{n}_i\hat{n}_j$ is the projection tensor with $\hat{n}$ being the unit vector pointing from the detector to the source, and $\mathbb{I}^{ij}$ is the inertia tensor. In the center-of-mass  frame, we have
\be
\mathbb{I}^{ij}=I^{ij}+J^{ij},
\ee
with
\be
I^{ij}=\mu r^2n^in^j,
\ee
being the inertia tensor in the case without the tidal interaction
and
\be\label{tidalinertiatensor}
J^{ij}=2M^6 \bk q \frac{n^i n^j}{r^3},
\ee
being the inertia tensor induced by the tidal field.

In the original AK mode, the inertia tensor is decomposed as a sum of harmonics of the radial orbital frequency
$I^{ij}=\sum_n I_n^{ij}$,
with
\bea
a_n^{(0)}&=&\frac{1}{2}(\ddot{I}^{11}_n-\ddot{I}^{22}_n),\\
b_n^{(0)}&=&\ddot{I}^{12}_n,\\
c_n^{(0)}&=&\frac{1}{2}(\ddot{I}^{11}_n+\ddot{I}^{22}_n),
\eea
where
\bea
a_n^{(0)}&=&\frac{n}{2}\mu (2\pi  M\nu)^{2/3}[(e^2-2)J_{n-2}(ne)+2eJ_{n-1}(ne)\nonumber\\
&&-2eJ_{n+1}(ne)+(2-e^2)J_{n+2}(ne)]\cos[n\Phi(t)],\nonumber\\
\eea
\bea
b_n^{(0)}&=&-\mu n (2\pi M\nu)^{2/3}(1-e^2)^{1/2}[J_{n-1}(ne)\\
&&-e(J_{n+1}(ne)+J_{n-2}(ne)+J_{n+2}(ne))] \sin[n\Phi(t)],\nonumber
\eea
\bea
c_n^{(0)}&=&-\frac{n}{2}e\mu  (2\pi M\nu)^{2/3} (e J_{n-2}(ne)-2J_{n-1}(ne)\nonumber\\
&&+2J_{n+1}(ne)-e J_{n+2}(ne)) \cos[n\Phi(t)].
\eea
where $J_n$ are Bessel functions of the first kind. The detailed derivation of above formulae can be found in \cite{Maggiore:2007ulw} and one can check that above expressions are equivalent to the ones in \cite{Peters:1963ux}.

Similarly, the tidal-induced inertia tensor can also be decomposed as $J^{ij}=\sum_n J_n^{ij}$, with
\bea
a_n^{(\rm{T})}&=&\frac{1}{2}(\ddot{J}^{11}_n-\ddot{J}^{22}_n),\\
\label{anT}
b_n^{(\rm{T})}&=&\ddot{J}^{12}_n,\label{bnT}\\
c_n^{(\rm{T})}&=&\frac{1}{2}(\ddot{J}^{11}_n+\ddot{J}^{22}_n).\label{cnT}
\eea
where
\begin{eqnarray}
  a_n^{(\rm{T})}
  & = &  \bk  \left( - n^2  \mu \right)(2\pi M\nu)^4 (X_n^{- 3, 2} + X_{- n}^{- 3, 0})   \cos [n \Phi],\nonumber\\
  b_n^{(\rm{T})}  & = &  \bk  \left( - n^2  \mu \right)(2\pi M\nu)^4 (X_n^{- 3, 2} - X_{- n}^{- 3, 2})  \sin [n \Phi],\nonumber\\
  c_n^{(\rm{T})} & = &  \bk  \left( - n^2  \mu \right) (2\pi M\nu)^4  (X_n^{- 3, 0} +
  X_{- n}^{- 3, 0}) \cos [n \Phi].\nonumber\\
\end{eqnarray}
The complete derivation of these expressions is lengthy so is presented in the Appendix \ref{appA}. Here $X_k^{ij}$
are Hansen coefficients \cite{Breiter:2004} which are useful in celestial mechanics when handling the Fourier decomposition involving Keplerian orbits, e.g., \cite{Mikoczi:2015ewa}. The Hansen coefficients can be expressed in terms of Bessel function series (see Eq. (\ref{Hansencoeff}) for explicit expressions) and the related ones appearing in above formulae truncated at finite orders are given by
\begin{eqnarray}
  &&X_k^{- 3, 0}\frac{(1 - \beta^2)^3}{(1 + \beta^2)^2} \nonumber\\
 && =  (1 + \beta^2)
  J_k (k e) + 2 \beta^{} J_{k - 1} (k e) + \beta^2 (3 - \beta^2) J_{k - 2} (k
  e)\nonumber\\
  &  & +  2 \beta^3 (2 - \beta^2)^{}
  J_{k - 3} (k e) + \beta^4 (5 - 3 \beta^2) J_{k - 4} (k e)\nonumber\\
  && + 2 \beta^5 (3 - 2
  \beta^2) J_{k - 5} (k e) +  \beta^6 (7 - 5 \beta^2)
  J_{k - 6} (k e)\nonumber \\
&&+  2 \beta^7 (4 - 3 \beta^2) J_{k - 7} (k e)
  + \beta^8 (9 - 7
  \beta^2) J_{k - 8} (k e)\nonumber\\
  &  & +2 \beta^9 (5 - 4 \beta^2)
  J_{k - 9} (k e) + \beta^{10} (11 - 9 \beta^2) J_{k - 10} (k e) \nonumber\\
  &&+ 2
  \beta^{11} (6 - 5 \beta^2) J_{k - 11} (k e)+ \beta^{12} (13 - 11 \beta^2)
  J_{k - 12} (k e)\nonumber\\
  &  & + 2 \beta J_{k + 1} (k e) +
  \beta^2 (3 - \beta^2)^{} J_{k + 2} (k e)\nonumber \\
  &&+ 2 \beta^3 (2 - \beta^2) J_{k + 3}
  (k e),
\end{eqnarray}
\begin{eqnarray}
  X_{- k}^{- 3, 0} & = & X_k^{- 3, 0},
\end{eqnarray}
\begin{eqnarray}
  &&\frac{X_k^{- 3, 2}}{(1 + \beta^2)^2} \nonumber\\
  &&=  J_{k - 2} (k e) + 4 \beta J_{k - 3} (k e)
  + 10 \beta^2 J_{k - 4} (k e)\nonumber\\
  &  & + 20 \beta^3 J_{k - 5} (k e) + 35 \beta^4 J_{k - 6} (k
  e) + 56 \beta^5 J_{k - 7} (k e)\nonumber\\
  &  & +84 \beta^6 J_{k - 8} (k e) + 120 \beta^7 J_{k - 9} (k
  e) + 165 \beta^8 J_{k - 10} (k e)\nonumber\\
  &  & +220 \beta^9 J_{k - 11} (k e) + 286 \beta^{10} J_{k -
  12} (k e) \nonumber\\
  &&+ 364 \beta^{11} J_{k - 13} (k e) + 455 \beta^{12} J_{k - 14} (k e)\nonumber\\
  && + 560 \beta^{13} J_{k
  - 15} (k e)+680\beta^{14}J_{- k - 16}(ke)\nonumber\\
  &&+816\beta^{15}J_{- k - 17}(ke),
\end{eqnarray}

\begin{eqnarray}
  &&\frac{X_{- k}^{- 3, 2}}{(1 + \beta^2)^2 }\nonumber \\
  &&= J_{- k - 2} (- k e) + 4 \beta J_{- k
  - 3} (- k e) + 10 \beta^2 J_{- k - 4} (- k e)\nonumber\\
  &  & + 20 \beta^3 J_{- k - 5} (- k e) + 35 \beta^4 J_{- k -
  6} (- k e) \nonumber\\
  &&+ 56 \beta^5 J_{- k - 7} (- k e) + 84 \beta^6 J_{- k - 8} (- k e) \nonumber\\
  &&+ 120 \beta^7 J_{- k -
  9} (- k e) + 165 \beta^8 J_{- k - 10} (- k e)\nonumber\\
  &  & + 220 \beta^9 J_{- k - 11} (- k e) + 286 \beta^{10}
  J_{- k - 12} (- k e) \nonumber\\
  &&+ 364 \beta^{11} J_{- k - 13} (- k e)+  455 \beta^{12} J_{- k - 14} (- k e)
  \nonumber\\
  && + 560 \beta^{13}
  J_{- k - 15} (- k e),
\end{eqnarray}
where
\be\label{betae}
\beta = \frac{\left( 1 - \sqrt{1 - e^2} \right)}{e} .
\ee
We have confirmed that the above formulas yield a relative error of less than $0.1\%$ when compared to the exact values of the Hansen coefficients for $e\leq0.75$ and $k\leq20$. To see this, let us define the relative error as 
\be
\epsilon^{ij}_k=\left|\frac{X^{ij}_k|_{\rm{exa}}-X^{ij}_k|_{\rm{app}}}{X^{ij}_k|_{\rm{exa}}}\right|,
\ee
where $X^{ij}_k|_{\rm{exa}}$ means the exact value of the Hansen coefficient and $X^{ij}_k|_{\rm{app}}$ denotes the above expression truncated at finite orders. As shown in Table \ref{relaeerrorHansen}, for a given $e=0.75$, the relative error is always smaller than $10^{-3}$ for $k\leq20$. Particularly, the relative error of $X^{-3,2}_{-k}$ is always smaller $10^{-10}$ in this case.
\begin{table*}[!htbp]
	\caption{The relative error between the exact  and approximate values of the Hansen coefficients are listed, where the eccentricity is taken as $0.75$.
	}\label{relaeerrorHansen}
	\begin{center}
		\setlength{\tabcolsep}{2mm}
		\begin{tabular}{|c|c|c|c|c|c|c|c|}
			\cline{1-6}
			$k$	& $1$ &$5$ & $10$ & $15$ &$20$ 
			\\
			\hline
			$\epsilon^{-3,0}_k$	&$0.1\times10^{-4}$ &$1.5\times10^{-4}$
			&$4.4\times10^{-4}$  &$2.3\times10^{-4}$ &
			$6.7\times10^{-4}$               	
			\\
			\hline	
				$\epsilon^{-3,2}_k$	&$<10^{-9}$ &$1.7\times10^{-9}$
			&$2.4\times10^{-4}$  &$3\times10^{-4}$ &
			$2.4\times10^{-4}$ 
				\\
			\hline	
			$\epsilon^{-3,2}_{-k}$	&$<10^{-15}$ &$5.7\times10^{-15}$
			&$0.14\times10^{-11}$  &$1.8\times10^{-11}$ &
			$8.4\times10^{-11}$ 
			\\
			\hline	
		\end{tabular}
	\end{center}
\end{table*}
When $e$ is small, the above formulae behave not very well at large $k$ but very well at small $k$.  As shown in Table \ref{relaeerrorHansensmalle}, for  $X^{-3,0}_{k}$ with $e=0.1$, the Hansen coefficients with $k>10$ can be safely neglected since they are too small to be effective. In this case, the Hansen coefficients with lower $k$ are  dominant and the ones with higher $k$ become irrelevant. Therefore, from a practical point of view, the above formulae work very well in the small $e$ case. 

\begin{table*}[!htbp]
	\caption{The relative error and the exact value of $X^{-3,0}_{k}$ with $e=0.1$.
	}\label{relaeerrorHansensmalle}
	\begin{center}
		\setlength{\tabcolsep}{2mm}
		\begin{tabular}{|c|c|c|c|c|c|c|c|}
			\cline{1-6}
			$k$	& $1$ &$5$ & $10$ & $15$ &$20$ 
			\\
			\hline
			$\epsilon^{-3,0}_k$	&$5.5\times10^{-13}$ &$4.8\times10^{-12}$
			&$8.6\times10^{-10}$  &$2.3\times10^{-14}$ &
			$4.6\times10^{-3}$               	
			\\
			\hline	
			$X^{-3,0}_{k}$	&$0.15$ &$6.9\times10^{-5}$
			&$3.9\times10^{-9}$  &$2.0\times10^{-13}$ &
			$1.0\times10^{-17}$ 
			\\
			\hline	
		\end{tabular}
	\end{center}
\end{table*}

Using above harmonic decomposition of the inertia tensor, we can express the GW strain at the detector position as a sum of harmonics of the radial orbital frequency as well. The GW strain at the detector can be decomposed as 
\be
h_{ij}(t)=A^+(t)H_{ij}^+(t)+A^\times(t) H^\times_{ij}(t),
\ee
where $H_{ij}^+$ and $H_{ij}^\times$ are the two polarization basis tensors constructed with the unit vector pointing from the detector to the source $\hat{n}$ and  the unit vector $\hat{L}$,
\be
H_{ij}^+(t)=\hat{p}_i\hat{p}_j-\hat{q}_i\hat{q}_j,\quad H_{ij}^\times(t)=\hat{p}_i\hat{q}_j+\hat{q}_i\hat{p}_j,
\ee
with
\be
\hat{p}=\frac{\hat{n}\times\hat{L}}{|\hat{n}\times\hat{L}|},\quad \hat{q}=\hat{p}\times\hat{n},
\ee
and $A^+$ and $A^\times$ are the amplitudes of the two polarizations. The amplitudes of the two polarisations can be further expressed as $n$-harmonics of the radial orbital frequency as well, i.e., $A^+  \equiv \frac{1}{D} \sum_n A_n^+$ and 
$A^\times \equiv   \sum_n A_n^\times $, with
\bea
 A_n^+  &=&  -\Big[1 + (\hat{L} \cdot \hat{n})^2\Big]\Big[a_n \cos2\gamma -b_n\sin2\gamma\Big]
 \nonumber\\
&& +  c_n\Big[1-(\hat{L} \cdot \hat{n})^2\Big], 
	\eea
\be
A_n^\times =2 (\hat{L} \cdot \hat{n}) \Big[b_n\cos2\gamma+a_n\sin2\gamma\Big],
\ee
where in the presence of the tidal interaction we have
\bea
a_n&=&a_n^{(0)}+a_n^{(\rm{T})},\\
b_n&=&b_n^{(0)}+b_n^{(\rm{T})},\\
c_n&=&c_n^{(0)}+c_n^{(\rm{T})}.
\eea
In above expressions, $\gamma$ is an azimuthal angle measuring the direction of pericentre with respect to the orthogonal projection of $\hat{n}$ onto the orbital plane, which further depends on $\tilde{\gamma}$ and $\alpha$ (see \cite{Barack:2003fp} for more details). 
So far we have seen how the relevant parameters of the orbital evolution equaitons enter into the GW strain. In fact, if we neglect the spin of the CO, an EMRI event can be completely characterized by 14 degrees of freedom. However, in the present case, an additional parameter, namely the TLN, must be included. These parameters are listed as follows:
\bea
\Big\{&&m, M, a,e_0, \tilde{\gamma}_0,\Phi_0,\lambda,\bk,\nonumber\\
&&\cos\theta_S, \phi_S,\alpha_0, 
\cos\theta_K,\phi_K, D,t_0 \Big\}.
\eea
Here, $t_0$ is a time parameter  at which the radial orbital frequency equates some fiducial  frequency $\nu_0$. Since the orbital evolution equations are solved in the reverse time direction, all quantities with subscript $0$ can be understood as initial values. Moreover, the angles $(\theta_S, \phi_S)$ are the direction to the source and $(\theta_K, \phi_K)$ represent the direction of the SMCO's spin. The first eight parameters are instrinsic  \cite{Buonanno:2002ft} , in the sense that they describe the system without reference to the location or orientation
of the observer. In contrast, the remaining seven are extrinsic parameters. 

To perform data analysis, we need to know the detector's response to the GW signal. Since the equilateral triangle detectors such as LISA  can
be used to construct two independent Michelson interferometers, the signal responded by such two interferometers can be decomposed into $n$-harmonics as well, so 
\be
h_{I,II}=\frac{\sqrt{3}}{2}\left(F^+_{I,II}A^++F^\times_{I,II}A^\times\right),
\ee
where $F^{+,\times}_{I,II}$ are antenna pattern function of the detector \cite{Cutler:1994ys}.
\section{Results}\label{results}
In this section, we will first introduce the method of analyzing the GW waveforms and evaluating the measurement of the tidal deformability of the SMCO using the future space-based interferometer LISA. Then we show the explicit results of the comparison of the two kind waveforms with and without the tidal interaction, and the constraint on the TLN of the SMCO for events detectable by LISA.
\subsection{Method of GW analysis}
To assess the strength of the effect of the tidal deformability of the SMCO on the EMRI waveforms to be measurable by a space-based GW detector, it is convenient to introduce the overlap
$\mathcal{O}$ between two waveforms $h_1(t)$ and $h_2(t)$,
\begin{equation}
\mathcal{O}(h_1|h_2)=\frac{<h_1|h_2>}{\sqrt{<h_1|h_1><h_2|h_2>}},
\end{equation}
where the noise-weighted inner product $<h_1|h_2>$ is defined by
\begin{align}\label{inner}
<h_1 |h_2 > =2\int^\infty_0 df \frac{\tilde{h}_1^*(f)\tilde{h}_2(f)+\tilde{h}_1(f)\tilde{h}_2^*(f)}{S_n(f)},
\end{align}
where the quantities with tilde stand for the Fourier transform, the star means complex conjugation,  and $S_n(f)$
is noise power spectral density of a space-borne GW detector, such as LISA~\cite{LISA:2017pwj}. The explicit expression of $S_n(f)$ for LISA is presented in the Appendix \ref{appC}.
It is more often use the mismatch $\mathcal{M}$ to quantify the difference between two  waveforms, with the definition given by
\begin{equation}
    \mathcal{M} \equiv 1- \mathcal{O}(h_1|h_2).
\end{equation}
If the two waveforms are identical, then the overlap
between them equates unity and so their mismatch is
zero.  A criterion to distinguish two waveforms by a GW detector is that their
mismatch  has to be larger than $\mathcal{D}/2\rho^2$  \cite{Flanagan:1997kp,Lindblom:2008cm}, where $\rho$ is the signal-to-noise ratio (SNR) of the detected signal and $\mathcal{D}$ denotes the number of the intrinsic parameters  of an EMRI system. Including the TLN of the SMCO, there are eight intrinsic parameters in the present case.
The SNR threshold for EMRI that can be detected is usually chosen to be $20$ \cite{Babak:2017tow}. Then  two waveforms with mismatch larger than
0.01 can be resolved by space-based detectors.

To quantify the capability of space-based GW detectors  to constrain the
 parameters of the EMRIs, we adopt the fisher informational matrix (FIM) method  \cite{Vallisneri:2007ev}.
In the high SNR limit, the FIM can capture the lowest-order expansion of the posteriors.
The FIM is defined by
\begin{equation}
    \Gamma_{ab}=\Big( \frac{\partial h}{\partial \lambda_a} \Big| \frac{\partial h}{\partial \lambda_b} \Big),
\end{equation}
where $\lambda_a$, $a=1,2,...,$ are the parameters appearing in the waveform and  the inner product $(|)$ is  defined by Eq. \eqref{inner}.
When the SNR of the GW signal is large, the variance-covariance matrix can be
obtained as the inverse of the FIM
\be
\Sigma_{ab}\equiv <\Delta \lambda_a \Delta\lambda_b>=(\Gamma^{-1})_{ab}.
\ee
From the variance-covariance matrix, the uncertainty of the $a$-th parameter $\lambda_a$ can be
obtained as
\begin{eqnarray}
\delta\lambda_a=\Sigma_{aa}^{1/2}.
\end{eqnarray}
Note that the applicability of the FIM method requires
the linear signal approximation to be valid. For EMRI events with SNR $\rho=20$ detected by LISA, the FIM is adoptable, which has been illustrated in \cite{Zi:2022hcc}. Moreover, the numerical stability of the inverse FIM is also required. This is discussed in Appendix \ref{appB}.
\subsection{Waveforms and mismatch}

\begin{figure*}[th]
\centering
\includegraphics[width=0.45\textwidth]{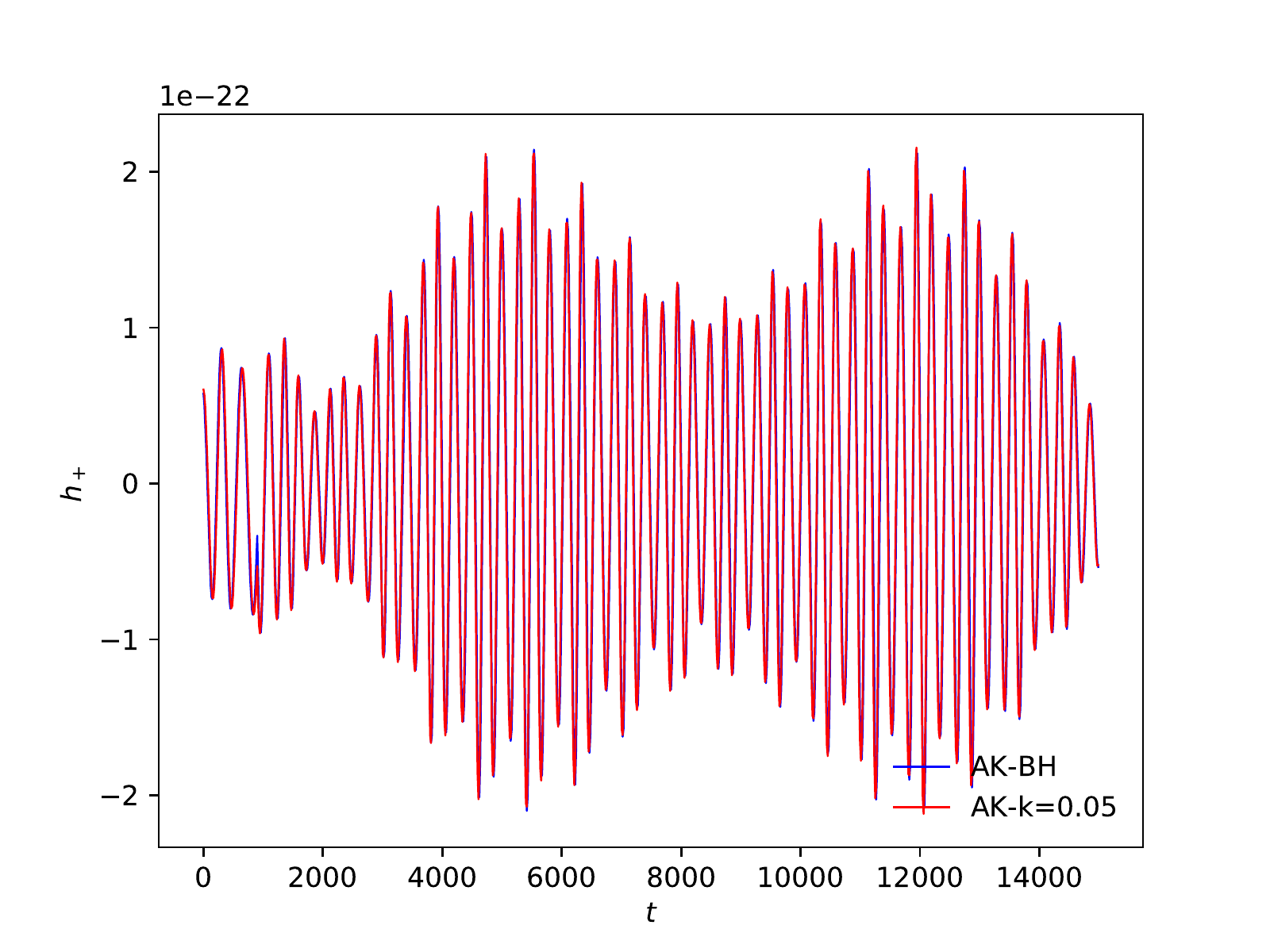}
    \includegraphics[width=0.45\textwidth]{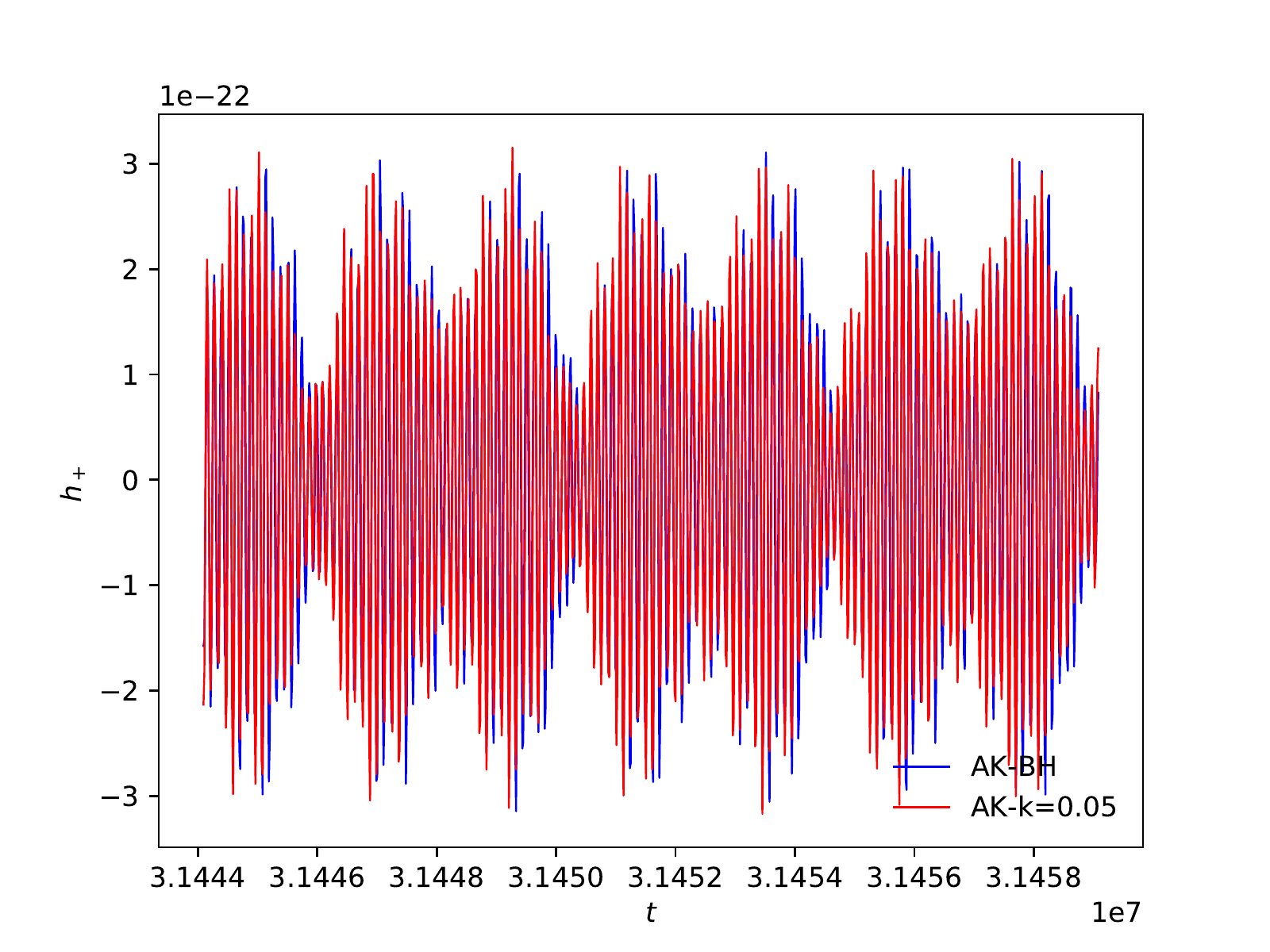}
	\caption{Comparison among plus polarization $h_+$ of AK waveforms from  EMRIs in the case of spin $a=0.8$ for $\bk=0$ and $0.05$, where the initial frequency is set as $\nu_0=1 \textup{mHz}$. The length of the waveform is 1 year, and the left panels represent the waveform for the first 30000 seconds, while the right panels for the last 30000 seconds.
 }\label{waveformscomparison}
\end{figure*}
Solving the orbital evolution equations and plugging the time-varying orbital parameters into the expression of the GW strain at the detector, we can obtain the AK waveforms in the time domain numerically. In Fig. \ref{waveformscomparison} we show the plus polarization $h_+$ of the AK waveforms with and without the tidal interaction.  Since we are interested in the impact of the tidal deformability of the SMCO on the waveforms, we only let the TLN free and keep other parameters fixed as follows:
$t_0=1$ years, $D=1$ Gpc, $m=10~{\rm M}_\odot$, $M=10^{6}~{\rm M}_\odot$, $e_0=0.1$,
$\lambda=\pi/3$, $\tilde{\gamma}_0=5\pi/6$,
$\alpha_0=4\pi/5$, $\theta_S=\pi/5$,
$\phi_S=\pi/4$, $\theta_K=2\pi/3$, $\phi_K=3\pi/4$, $\Phi_0=\pi/3$, and $\nu_0=1 \text{mHz}$. To better illustrate the comparison of waveforms with and without tidal interaction, we will deviate from the original AK waveform procedure, where the orbital evolution equations were solved in the reverse time direction. Instead, we will solve the equations in the forward time direction. Therefore, in this context, $t_0$ represents the length of the waveforms, and all quantities with subscript $0$ indicate values at $t=0$, not $t=t_0$. From Fig.\ref{waveformscomparison} we can observe that the AK waveform is significantly affected by the tidal deformability of the SMCO.  Even the TLN is as small as $0.05$, the phase difference between the two waveforms becomes noticeable if the signal lasts for one year.

\begin{figure}[th]
	\centering
	\includegraphics[width=0.45\textwidth]{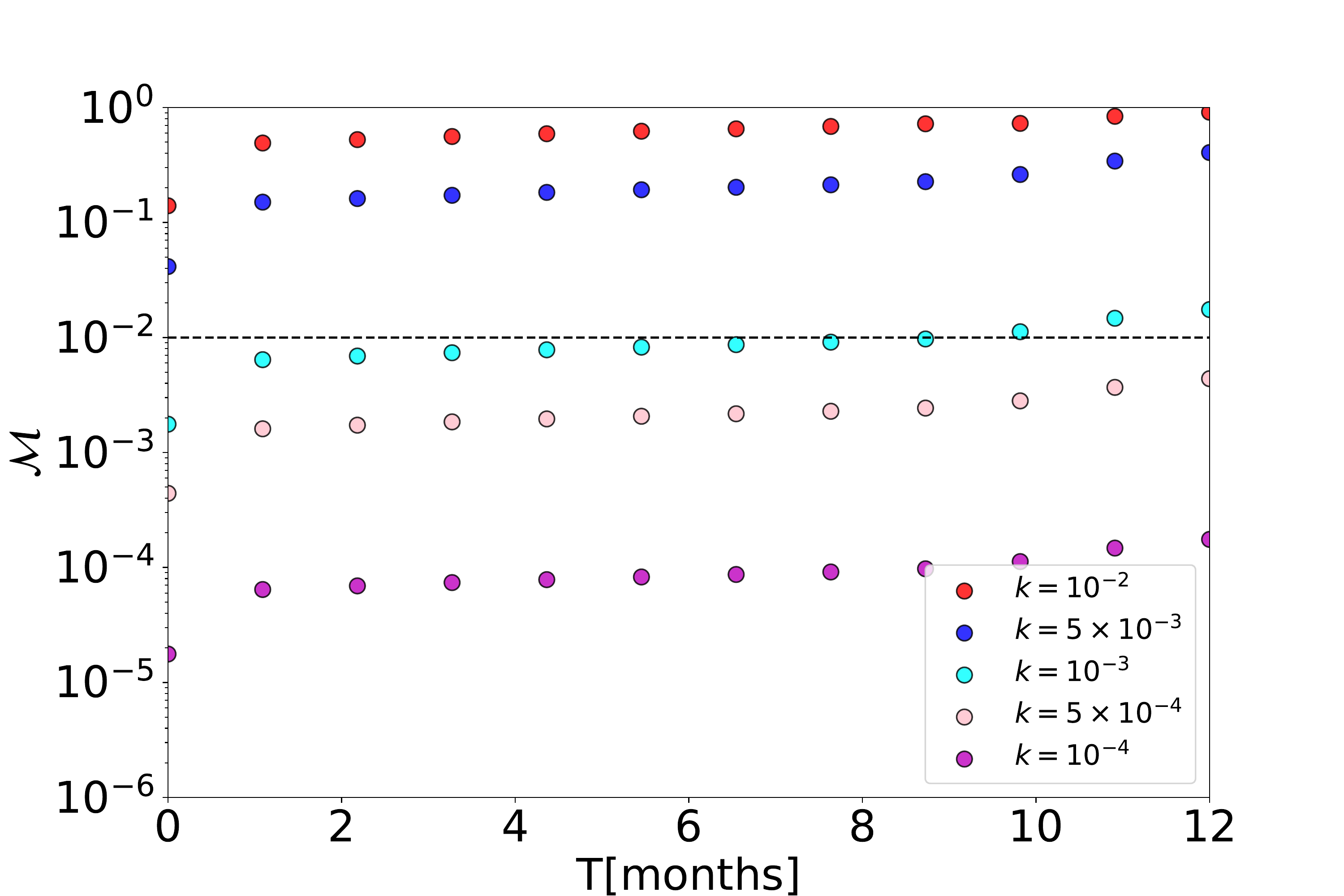}
	\caption{The mismatch $\mathcal{M}$ of different values of the TLN $\bk$ as a function of observation time for LISA is plotted, the dashed lines represent the threshold for SNR=20.
		The source parameters are set as $M=10^6M_\odot$, $a=0.4$,  
		the other parameters keep same with the previous configurations in Fig.~\ref{waveformscomparison}.
	}\label{mismatch:ObservationTime:LISA}
\end{figure}
To assess the imprint of tidal deformability of the SMCO on the EMRI waveforms quantitatively, we calculate the mismatches between the original AK waveform and  the ones with different values of the TLN $\bk$. 
As shown in Fig. \ref{mismatch:ObservationTime:LISA}, the mismatches as functions of the observation time are plotted. The source parameters are set as $M=10^6M_\odot$ and $a=0.4$. For 1 year observation of LISA and with SNR $\rho=20$, the mismatches can exceed the threshold value $\mathcal{M}_{\rm min} = 0.01$ as long as the TLN is $O(10^{-3})$. 

\begin{figure*}[th]
	\centering
	\includegraphics[width=0.45\textwidth]{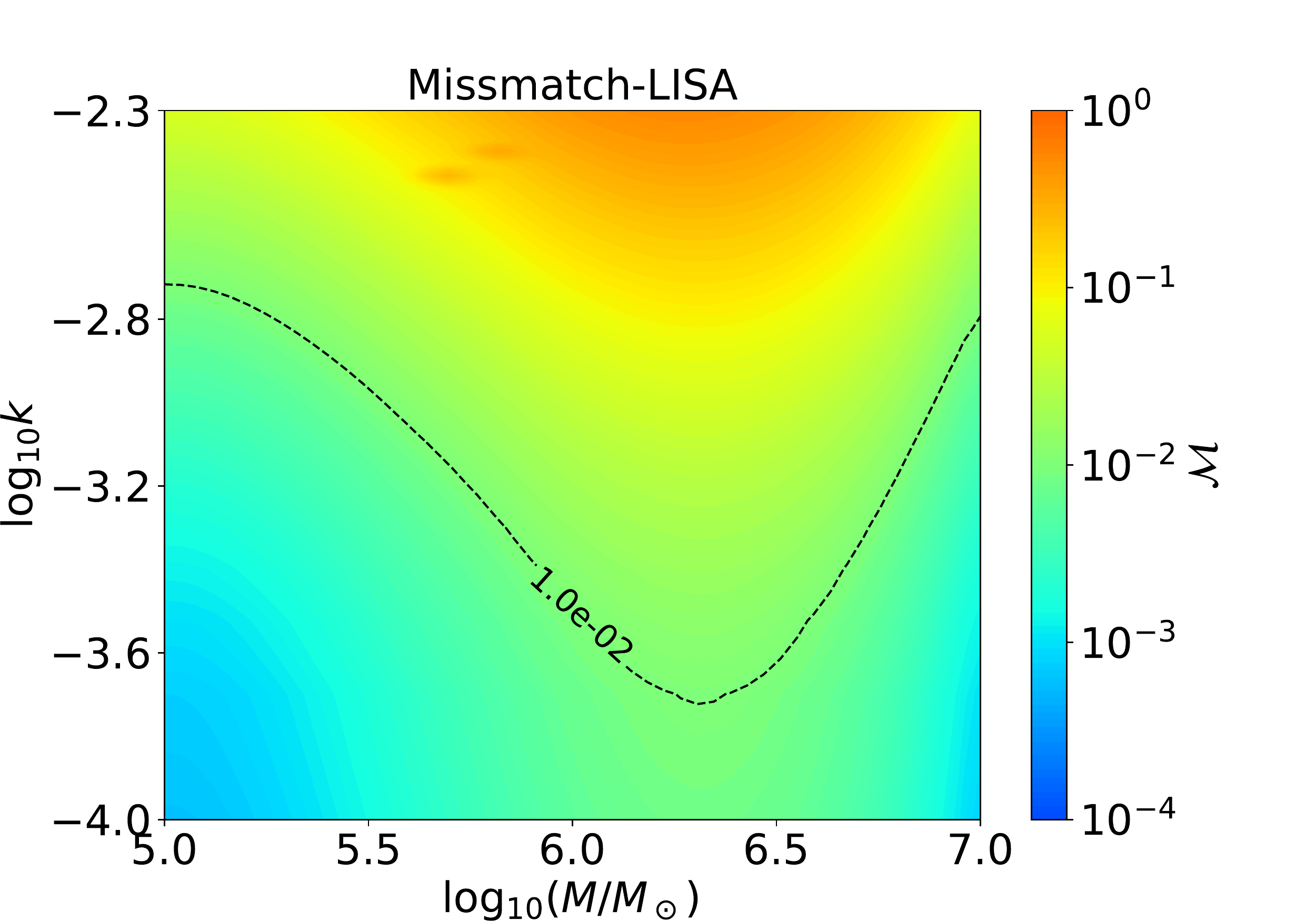}
	\includegraphics[width=0.45\textwidth]{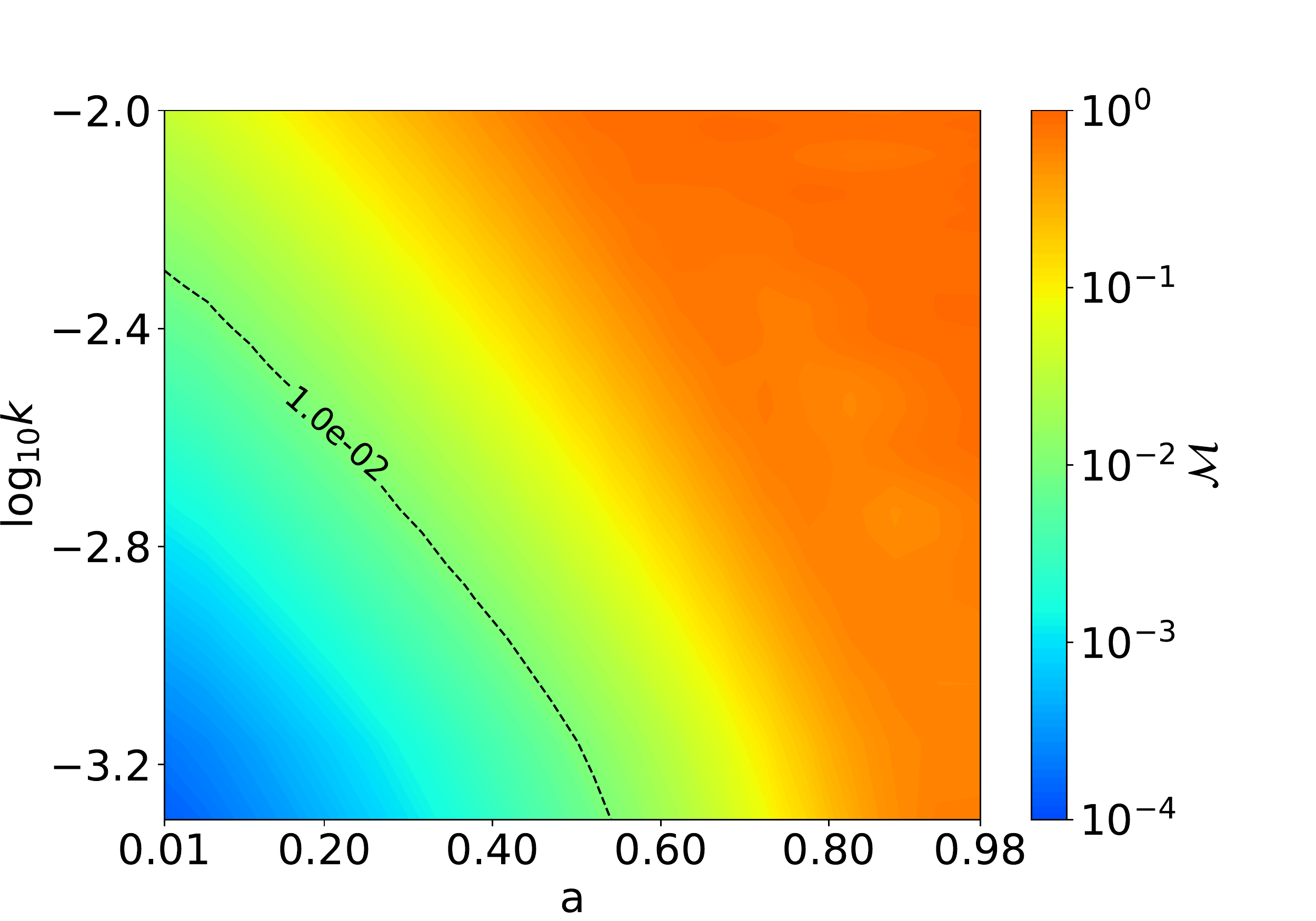}
	\caption{The contour plot of the mismatch $\mathcal{M}$ as functions of $\log_{10} {\bk}$ and  $\log_{10} {M}$ (left),  or $a$ (right) with respect to LISA. In the left panel $a=0.8$ and in the right panel $M=10^6 M_\odot$. The black dashed line denotes to the threshold value for SNR=20 and
		the other parameters keep same with the previous configurations in Fig.~\ref{waveformscomparison}.
	}\label{mismatchContourf}
\end{figure*}

To further study the impacts of the mass and the spin of the SMCO on the mismatches, in Fig.  \ref{mismatchContourf} we plot  the mismatch as functions of $\bk$ and $M$ or $a$.The black dotted lines represent the contour of mismatch equal to the threshold $\mathcal{M}=0.01$, it indicates that LISA can distinguish whether the SMCO in an EMRI has $\bk\neq0$ if the system is located beyond this curve. We can see that for SMCO masses close to $10^{6.5} M_\odot$, the TLN of SMCO that can be resolved by LISA is the smallest. This indicates that the mass of SMCO has a significant impact on the TLN detected by the LISA detector. Moreover, the value of the TLN on the threshold line decreases with the spin of the SMCO. When the spin is larger than $0.6$, the TLN that can be resolved by LISA is smaller than $10^{-3}$. Therefore, under  suitable  scenarios, the LISA is able to distinguish SMCO with TLN as small as $10^{-3}$.

\begin{figure*}[th]
	\centering
	\includegraphics[width=0.45\textwidth]{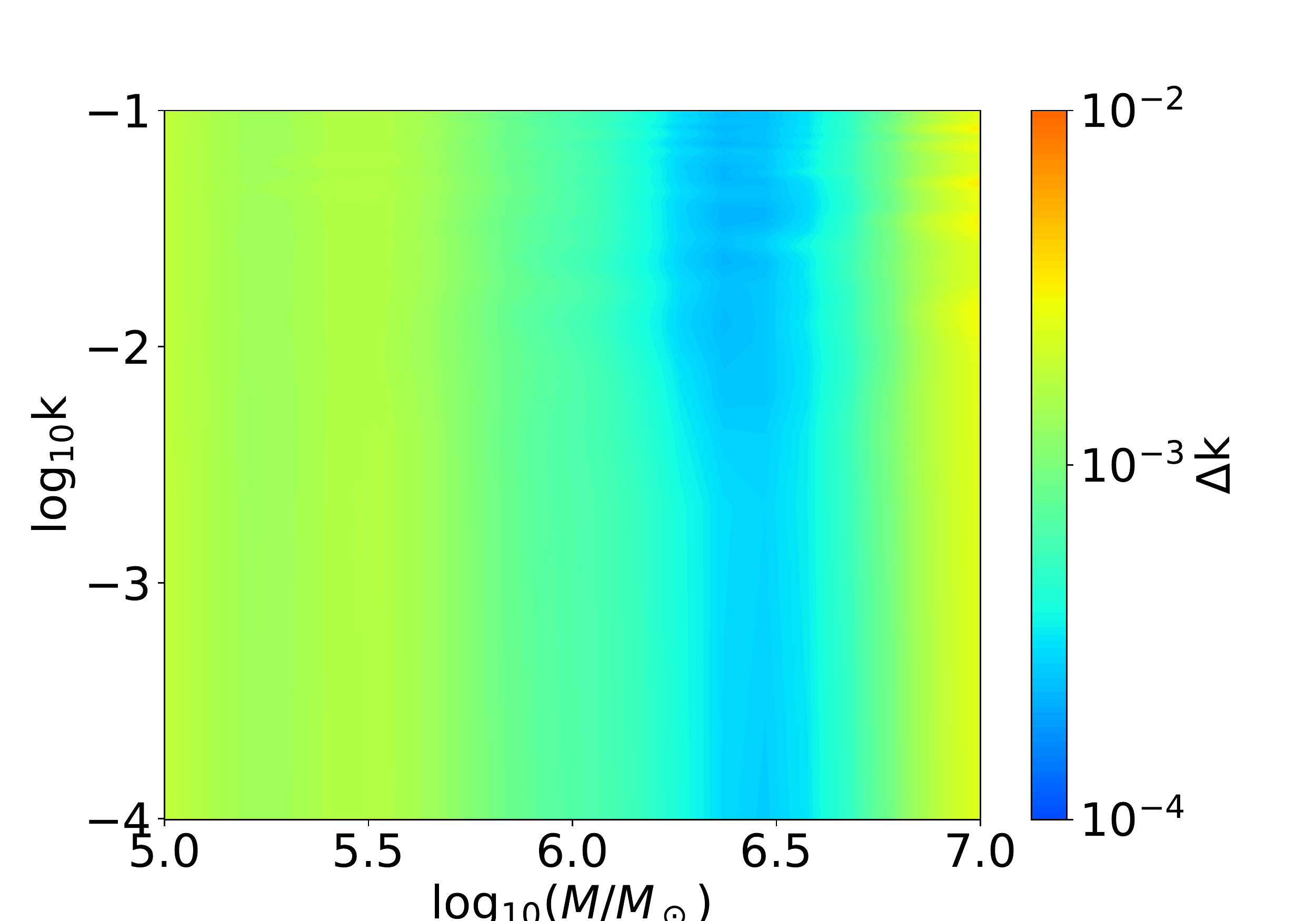}
	\includegraphics[width=0.45\textwidth]{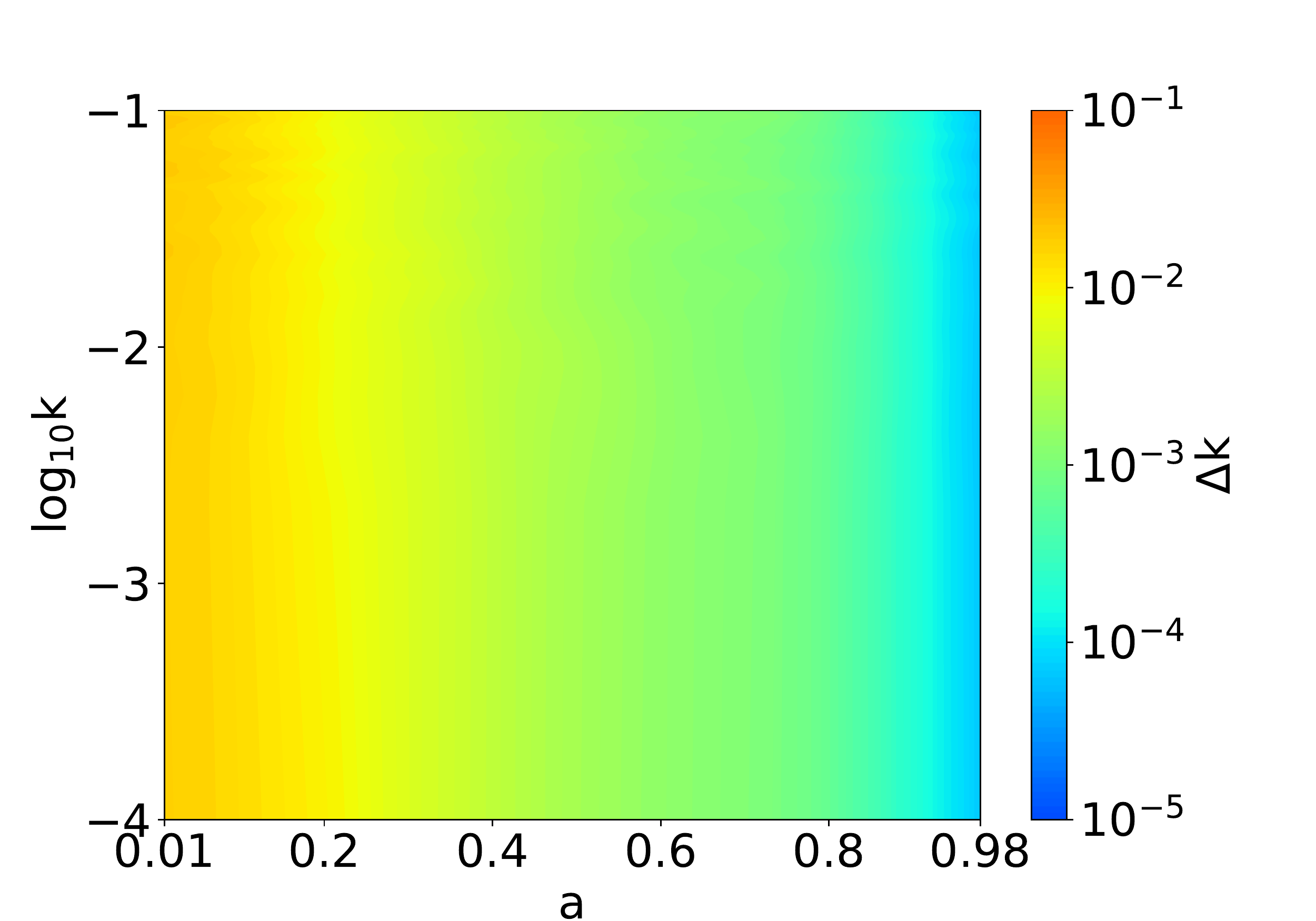}
	\caption{The contour plot of the parameter estimation accuracy as functions of $\log_{10} {\bk}$ and  $\log_{10} {M}$ (left),  or $a$ (right) with respect to LISA. In the left panel $a=0.8$ and in the right panel $M=10^6 M_\odot$. The 
		other parameters keep same with the previous configurations in Fig.~\ref{waveformscomparison}.
	}\label{parameterestimation}
\end{figure*}
\subsection{Constraint on TLN}
In this subsection, we perform the parameter estimation
for the TLN  using the FIM method. In the original AK waveform, the cutoff for the inspiral is determined by the last stable orbit of a Schwarzschild or Kerr BH. However, in the present case, the length of the waveforms is fixed to 1 year to avoid the unknown effects of the tidal interaction on the cutoff. By taking the central values of the TLN to
a given value, we can study the effects of various parameters  on the constraints for the TLN. Here we only focus on the effects from the mass $M$, the spin
parameter $a$ and  the TLN $\bk$.

As depicted in Fig. \ref{parameterestimation}, when the spin parameter is fixed at $a=0.8$, the impact of the SMCO mass on the uncertainty of the TLN is not a monotonous function. Interestingly, we observe that the most stringent constraint on the TLN can be achieved when the SMCO mass is close to $10^{6.5} M_\odot$, with a potential resolution of $10^{-4}$. This is because the EMRI system with a more massive SMCO produces GWs with lower frequencies. The sensitivity of the GW detector is closely tied to the GW frequency and, additionally, to the mass of the SMCO, as indicated by the sensitivity curve. Moreover, 
for a fixed mass $M=10^6M_\odot$, the uncertainty of the TLN decreases with the spin parameter $a$, so the SMCO with largest spin has the best constraint on the TLN. This is consistent with the study in \cite{Piovano:2022ojl}.  We can find that when $a>0.8$, the constraint on TLN can reach the level of $10^{-4}$. From both panels and Table \ref{TLNandconstraint}, we can see that the effects of the TLN values on the uncertainty of the TLN are not prominent. The reason for this phenomenon could be attributed to the fact that the phase of the waveform is depended linearly on the TLN. The calculation of the FIM involves the derivation of the waveform with respect to the TLN. As a consequence, the effect of the TLN may disappear in the constraint of itself.

\begin{table*}[!htbp]
	\caption{Constraints on different tidal love numbers $\bk$ of SMCO with mass $M=10^{6}M_\odot$ and spin $a=0.8$ are listed.
	}\label{TLNandconstraint}
	\begin{center}
		\setlength{\tabcolsep}{4mm}
		\begin{tabular}{|c|c|c|c|c|c|c|c|c|c|c|c|c|}
			\cline{1-10}
		$\bk$	& $10^{-4}$ &$10^{-3}$ & $10^{-2}$ & $0.05$ &$0.1$ &$0.5$ &$1.0$ & $5$ & $10$
			\\
			\hline
		$\Delta \bk/10^{-4}$	&$6.66$ &$6.51$
			 &$5.99$  &$6.09$ &
			$6.29$     &$7.87$ &$7.91$
			&$8.94$			&$9.87$             	
			\\
			\hline	
		\end{tabular}
	\end{center}
\end{table*}

\section{Summary}\label{summary}
In this paper, we investigated the effect of tidal deformability of a SMCO in an EMRI on the gravitational waveforms. Our study was carried out within the framework of the AK waveforms. Firstly, as the tidal interaction between the SMCO and the CO is proportional to the mass ratio, the known results of perturbed Keplerian orbits, obtained using the method of osculating orbital elements \cite{Poisson:2014book}, can be naturally applied in this scenario. Given that the mass ratio is very small, the conservative dynamics of the EMRI remain unaffected by tidal interaction up to leading order of the mass ratio. Consequently, the orbits can be approximated as Keplerian orbits, with the effect of tidal interaction being encoded only in the induced quadrupole moment.

We further calculated the energy and angular momentum fluxes using the quadrupole formulas in the presence of the tidal interaction. Then we derived the leading order equations describing the evolution of the radial orbital frequency and the eccentricity. On the other hand, the other orbital evolution equations in the AK model remain unchanged. Combine these leading order corrected equations with those higher-order PN equations
in the original AK model, the complete orbital evolution equations were obtained. Moreover, to express the GW strain as a sum of the harmonics of the radial orbital frequency, as was done in the original AK model. We used the Hansen coefficients method to perform the Fourier decomposition of the tidal-induced inertia tensor.

We found that the tidal deformability of the SMCO has a prominent effect on the AK waveforms. 
By calculating the mismatches between the AK waveforms with and without the tidal interaction, we showed that LISA can detect the deformed SMCO even if the parameter TLN is as small as $10^{-3}$, with just one year of observation. We then performed the parameter estimation precision for the TLN and found
that with one year observation LISA can measure them with accuracy to the level of $10^{-4}$ under suitable scenarios. 

In this paper, the tidal interaction was investigated in the post-Newtonian framework, so the results are not accurate in the strong-field regime. It would be intriguing to explore in the full relativistic regime to derive more compelling conclusions regarding the limits on the tidal deformability of the SMCO through the observations of  space-based GW detectors. On the other hand, there are many more interesting tidal effects can be explored using the EMRI GWs. For example, as discussed in \cite{Poisson:2014book}, the Newtonian tidal interaction also has the dissipative effect on the dynamics due to the presence of viscosity in the SMCO. The tidal dissipation introduces an additional perturbing force in the orbital equation of motion and is proportional to the mass ratio, thus can be handled with the method of the osculating orbital elements. The special case of circular orbits discussed \cite{Poisson:2014book} showed that the tidal dissipation indeed affects the orbital element  after the average over the orbital period has been performed. Besides, the explicit dependence of the waveform on the tidal interaction could be used to explore the properties of the environment around the central BH in an EMRI. This is because the environment around a BH could also give a non-zero TLN, see e.g. \cite{DeLuca:2021ite,DeLuca:2022xlz}. Moreover, the tidal field of a nearby astrophysical object or dark matter distribution of the EMRI could modify the orbital motion and induce an interesting phenomenon named  tidal resonances \cite{Bonga:2019ycj,Gupta:2021cno,Bronicki:2022eqa,Gupta:2022fbe}. This occurs when the linear combination of the fundamental frequencies of the orbits are commensurate. All these tidal effects must be considered in order to unravel the physics derived from the observations of the EMRI GWs. 

\begin{acknowledgments}
We are grateful to Jiandong Zhang for helpful discussion. The work is in part supported by NSFC Grant
No.12205104 and the startup funding of South China University of
Technology. This project is supported by MOE Key Laboratory of TianQin Project, Sun
Yat-sen University.
\end{acknowledgments}
\appendix
\section{The Fourier decomposition of the tidal-induced inertia tensor}\label{appA}
It is known that the Hansen coefficients  are defined as the Fourier amplitudes in the series
\be
 \left( \frac{r}{a_r} \right)^{\gamma} e^{i m \psi} = \sum^{\infty}_{k = -
   \infty} X^{\gamma, m}_k e^{i k \Phi},
\ee
where $\psi$ is the true anomaly, $\Phi$ is means anomaly, and $r$, $a_r$ the radial
distance and semi-major axis. $a_r$ is related to the semi-latus rectum by $p=a_r(1-e^2)$.

There are various forms of  Hansen coefficients expressed in terms of Bessel function series. In the following, we refer to the one in \cite{Breiter:2004},
\be \label{Hansencoeff}
X_k^{\gamma, j} = \frac{1}{(1 + \beta^2)^{\gamma + 1}} \sum_{s = -
   \infty}^{\infty} E^{\gamma, j}_{k - s} J_s (k e),
   \ee
where $\beta $ is given by Eq. (\ref{betae})
and $J_s(z)$ is the Bessel function of the first kind. Moreover, for $p\geq j$
\bea
 E^{\gamma, j}_p &=& (- \beta)^{p - j} C^{\gamma - j + 1}_{p - j}\\
 &&\times {}_2F (- \gamma
   - j - 1, p - \gamma - 1; p - j + 1 ; \beta^2),\nonumber
\eea
and for $p<j$,
\be
E^{\gamma,j}_p=E^{\gamma,-j}_{-p},
\ee
where $C^n_k$ denotes the binomial coefficient $n!/k!(n-k)!$ and ${}_2F(a,b;c;d) $ is the hypergeometric function.

Firstly, setting $\gamma = - 3$ and $m =
0$, Eq. (\ref{Hansencoeff}) gives
\be
 r^{- 3} = \sum_{k = 0}^{\infty} R^T_k \cos k \Phi,
 \ee
where
\begin{eqnarray}
  R_0^T & = & a_r^{- 3} X_0^{- 3, 0},\\
  R^T_k & = & a_r^{- 3} (X_k^{- 3, 0} + X_{- k}^{- 3, 0}) .
\end{eqnarray}
Secondly, setting $\gamma=-3$ and $m=2$, the real part of the Eq. (\ref{Hansencoeff}) gives
\be
 r^{- 3} \cos 2 \psi = \sum_{k = 0}^{\infty} P_k \cos k \Phi,
 \ee
where
\begin{eqnarray}
  P_0 & = & a_r^{- 3} X_0^{- 3, 2},\\
  P_k & = & a_r^{- 3} (X_k^{- 3, 2} + X_{- k}^{- 3, 2}),
\end{eqnarray}
and the imaginary part of the Eq. (\ref{Hansencoeff}) gives
\be
r^{- 3} \sin 2 \psi = \sum_{k = 1}^{\infty} Q^T_k \sin k \Phi,
\ee
where
\be
 Q^T_k = a_r^{- 3} (X_k^{- 3, 2} - X_{- k}^{- 3, 2}) .
 \ee
From these results, we obtain directly
\begin{eqnarray}\label{rm3cos2}
 r^{-3}\cos^2 \psi  = \sum_{k = 0}^{\infty} A^T_k \cos k \Phi,
\end{eqnarray}
where
\begin{eqnarray}
  A_0^T & = & \frac{1}{2} (R_0^T + P_0^T)\nonumber\\
  & = & \frac{1}{2} a_r^{- 3} (1 - e^2)^{- 3 / 2},
\end{eqnarray}
and for $k>0$
\begin{eqnarray}
  A_k^T & = & \frac{1}{2} (R_k + P_k)\nonumber\\
  & = & \frac{1}{2} a_r^{- 3} (X_k^{- 3, 0} + X_{- k}^{- 3, 0})\nonumber\\
  && + \frac{1}{2}
  a_r^{- 3} (X_k^{- 3, 2} + X_{- k}^{- 3, 2}) .
\end{eqnarray}
Next, it is easy to find
\begin{eqnarray}\label{rm3sin2}
r^{-3}\sin^2 \psi =  \sum_{k = 0}^{\infty} B^T_k \cos k \Phi,
\end{eqnarray}
where
\begin{eqnarray}
  B^T_0  & = & \frac{1}{2} (R_0^T - P_0^T)\nonumber\\
  & = & \frac{1}{2} a_r^{- 3} (1 - e^2)^{- 3 / 2},
\end{eqnarray}
and for $k>0$
\begin{eqnarray}
  B^T_k  & = & \frac{1}{2} (R_k^T - P_k^T)\nonumber\\
  & = & \frac{1}{2} a^{- 3} (X_k^{- 3, 0} + X_{- k}^{- 3, 0})\nonumber\\
  &&- \frac{1}{2}
  a^{- 3} (X_k^{- 3, 2} + X_{- k}^{- 3, 2}) .
\end{eqnarray}
Moreover, we have
\begin{eqnarray}\label{rm3cossin}
 r^{- 3}\cos \psi \sin^{} \psi& = & \sum_{k = 1}^{\infty} C_k \sin k \Phi,
\end{eqnarray}
where
\begin{eqnarray}
  C_k^T&=&  \frac{1}{2} Q_k^T\nonumber\\
  &=& \frac{1}{2} a^{- 3} (X_k^{- 3, 2} - X_{- k}^{- 3, 2}) .
\end{eqnarray}
From Eq. (\ref{tidalinertiatensor}), we obtain the components of the tidal-induced inertia tensor
\bea
J^{11}&=&2M^6\bk q r^{-3}\cos^2\psi,\\
J^{12}&=&2M^6\bk q r^{-3}\cos\psi \sin\psi,\\
J^{12}&=&2M^6\bk q r^{-3}\sin^2\psi.
\eea
Then from Eqs.(\ref{rm3cos2}),(\ref{rm3sin2}) and (\ref{rm3cossin}), we can decompose above components into
a sum of harmonics of the radial orbital frequency and the results are just Eq.(\ref{anT}), (\ref{bnT}) and (\ref{cnT}). It is worth noting that we can perform the Fourier decomposition of the inertial tensor in the absence of tidal interaction by setting $\gamma=2$ and $m=0, 2$ using a similar procedure. It can be verified that the obtained results are identical to those reported in \cite{Barack:2003fp}.

\section{Stability of the Fisher matrix}\label{appB}
In this appendix we assess the stability of the covariance matrix for the EMRI signals by following the procedure in Ref. \cite{Piovano:2021iwv,Franciolini:2022tfm}. The basic idea is to observe the behavior of the covariance matrices  when small perturbations in the components in Fisher matrices are imposed. This is characterized quantitatively by 
\begin{equation}\label{fim:stab}
	\delta_{\rm stability} \equiv \mathbf{max}_{\rm ij} \left[\frac{((\Gamma+F)^{-1} - \Gamma^{-1})^{ij}}{(\Gamma^{-1})^{ij}}\right]
\end{equation}
with a deviation matrix $F^{ij}$, whose elements is a uniform distribution $U\in [a,b]$.
We calculate the stability of the Fisher matrix using Eq. \eqref{fim:stab},
the result is listed in the following table.
\begin{table*}[!htbp]
	\caption{$\delta_{\rm stability}$ for different spins  of the SMCO with mass $M=10^{6}M_\odot$  is listed.
	}\label{FIMstability}
	\begin{center}
		\setlength{\tabcolsep}{5mm}
		\begin{tabular}{|c|c|c|c|c|c|}
			\hline
			\multirow{2}{*}{$U$}& \multicolumn{5}{|c|}
			{spin $a$}\\
			\cline{2-6}
			& $0.1$ &$0.3$ &$0.5$ &$0.7$ & $0.9$ \\
			\hline
			$\in[-10^{-7},10^{-7}]$ & $0.049~~$ &$0.044$ &$0.049$
			&$0.048$  &$0.029$
			\\		
			\hline
			$\in[-10^{-9},10^{-9}]$ & $0.032~~$ &$0.031$ &$0.036$
			&$0.039$  &$0.014$
			\\
			\hline	
			
		\end{tabular}
	\end{center}
\end{table*}

\section{Sensitivity curve }\label{appC}
The sky-averaged detector sensitivity for LISA can be give by in \cite{LISA:2017pwj,Babak:2017tow}
\begin{eqnarray}
	S_n(f)&=&\frac{20}{3}\frac{4S_{n}^\mathrm{acc}(f)+2S_{n}^\mathrm{loc}+S_{n}^\mathrm{sn}+S_{n}^\mathrm{omn}}{L^2} \nonumber \\
	& & \times
	\left[1+\left(\frac{2Lf}{0.41 c}\right)^2\right],
	\label{eq:sens}
\end{eqnarray}
where $L=2.5\times 10^{9}m$ is the arm length among satellites, and the noise 
$S_{n}^\mathrm{acc}(f)$, $S_{n}^\mathrm{loc}$,
$S_{n}^\mathrm{sn}$ and $S_{n}^\mathrm{omn}$ result from the low-frequency acceleration, local interferometer noise,
shot noise and other measurement noise, respectively. 
They can be written as the following according to LISA Pathfinder~\cite{Armano:2016bkm}
\begin{eqnarray}
	S_{n}^\mathrm{acc}(f) & = & \left\{9 \times 10^{-30}+3.24 \times 10^{-28}\left[\left(\frac{3\times10^{-5}~\mathrm{Hz}}{f}\right)^{10} \right. \right. \nonumber \\
	&  & \left. \left. + \left(\frac{10^{-4}~\mathrm{Hz}}{f}\right)^{2}\right]\right\}\frac{1}{(2\pi f)^4}\,\mathrm{{m^2\,Hz}^{-1}},
\end{eqnarray}
and the other noise expression are of the following
\begin{equation}
	\begin{split}
		&S_{n}^\mathrm{sn}= 7.92\times10^{-23}~\mathrm{{m}^2\,{Hz}^{-1}},\\
		&S_{n}^\mathrm{omn}=4.00\times10^{-24}~\mathrm{{m}^2\,{Hz}^{-1}},\\
		&S_{n}^\mathrm{loc}= 2.89\times10^{-24}~\mathrm{{m}^2\,{Hz}^{-1}}.
	\end{split}
\end{equation}


\end{document}